\documentclass[pre, nofootinbib, floatfix, notitlepage]{revtex4-1}
\usepackage[utf8]{inputenc}

\usepackage{amssymb,amsthm,amsmath, amsfonts}
\usepackage{color, graphicx, enumerate}
\usepackage[font={scriptsize}]{caption}
\usepackage{verbatim} 

\renewcommand{\qedsymbol}{\rule{0.7em}{0.7em}}

\begin{document}
\author{Katarzyna Macieszczak}
\affiliation{University of Nottingham, School of Mathematical Sciences, University Park, NG7 2RD Nottingham, UK}
\title{Quantum Fisher Information: Variational principle \\and simple iterative algorithm for its efficient computation}
\begin{abstract}
We derive a new variational principle for the quantum Fisher information leading to a simple iterative alternating algorithm, the convergence of which is proved. The case of a fixed measurement, i.e. the classical Fisher information, is also discussed.
\end{abstract}
\maketitle

\section{Introduction}

The Fisher information is an important concept in statistics. The Cramer-Rao inequality states that for any~unbiased estimator $\widetilde{\varphi}$ of a parameter %$\varphi\in\mathbb{R}$ its variance $\Delta^2\widetilde{\varphi}$ is lower bounded by the inverse of the Fisher information $F_{\varphi}$:
$\varphi\in\mathbb{R}$ its variance $\Delta^2\widetilde{\varphi}$ is lower bounded by the inverse of the Fisher information:
\begin{equation}
\Delta^2\widetilde{\varphi}=\int_X \mathrm{d}x\, p_{\varphi}(x)\, (\widetilde{\varphi}(x)-\varphi)^2\geq F_{\varphi}^{-1},
\end{equation}
where the Fisher information  $F_{\varphi}$ is defined as follows:
\begin{equation}
F_{\varphi}=\int_{\{x\in X:\,p_{\varphi}(x)\neq 0 \}}  \mathrm{d}x\, p_{\varphi}(x)\,\left(\frac{\partial \log(p_{\varphi}(x))}{\partial \varphi}\right)^2.
\end{equation}

Let $x$ be a result of a POVM  measurement $\{\Pi_x\}_{x\in X}$  ($\Pi_x\in \mathcal{B}(\mathcal{H})$, $\Pi_x\geq 0$, $\Pi_x=\Pi_x^{\dagger}$,  $\int_X\mathrm{d}x\,\Pi_x=1$) which is performed on~a~quantum state on a Hilbert space $\mathcal{H}$ described by a density matrix $\rho_{\varphi}\in S=\{\rho\in \mathcal{B}(\mathcal{H}):\, \rho=\rho^{\dagger},\, \mathrm{Tr}\{\rho\}=1\}$, then $p_{\varphi}(x)=\mathrm{Tr}\{\rho_{\varphi}\Pi_x\}$. $F_{\rho_\varphi,\{\Pi_x\}}$ depends on the choice of measurement $\{\Pi_x\}_{x\in X}$, but whatever the measurement, we have:
\begin{equation}
F_{\rho_\varphi,\{\Pi_x\}}\leq F^Q_{\rho_\varphi}=\mathrm{Tr}\{\rho_\varphi L_{\varphi}^2\},\quad \frac{1}{2}\{L_{\varphi},\rho_{\varphi}\}=\frac{\mathrm{d}}{\mathrm{d}\varphi}\rho_{\varphi}, \label{eq:qFisher}
\end{equation}
where $F^Q_{\rho_\varphi}$ is the so called quantum Fisher information, $L_{\varphi}$ is the symmetric logarythmic derivative and $\{.\,,.\}$ is the~anticommutator. The eigenbasis of the $L_{\varphi}$ operator corresponds to the~optimal projective measurement for which the~inequality in Eq. (\ref{eq:qFisher}) is saturated. \\

Let us assume that $\rho_{\varphi}=e^{-i \varphi H}\rho \,e^{i \varphi H}$, where the generator $H$ belongs to the $L^2(\mathcal{H})$ space of the self-adjoint Hilbert-Schmidt operators on $\mathcal{H}$. The quantum Fisher information has the same value for all $\varphi$, therefore let us drop the index $\varphi$. We have $F_{\rho}^Q=\mathrm{Tr}\{\rho L_\rho^2\}$ and $\frac{1}{2}\{L_\rho,\rho\}=-i[H,\rho]$.  %\\

\section{Variational principle}

\begin{figure}[htb!]
\includegraphics[width=0.5\textwidth]{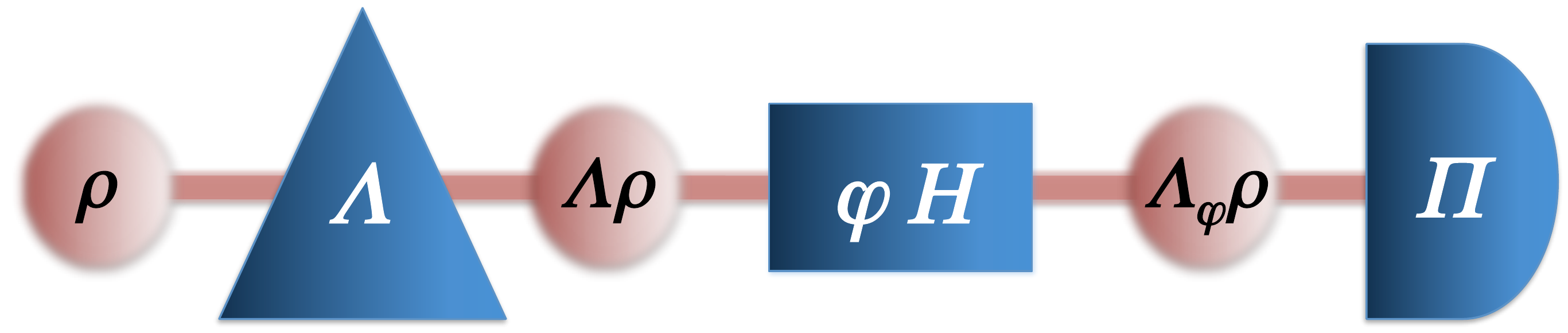}
\caption{The setup discussed in the draft: An input state $\rho$ first goes through a quantum channel $\Lambda$, then a parameter value $\varphi$ is imprinted using the $H$ generator and finally a POVM measurement $\{\Pi_x\}_{x\in X}$ is performed to retrive the information about the parameter. \label{fig:setup}} 
\end{figure}

%\textbf{Variational principle.}  
Let us consider the case in which the states used in the estimation are obtained as outputs from a quantum channel~$\Lambda$ (see Fig.~\ref{fig:setup}), i.e these are elements of $\Lambda(S)$ where $S$ is the set of density matrices.
Unless $\Lambda$ is a unitary channel, $\Lambda(S)$~is~restricted and~does~not~contain all density matrices, e.g. $\Lambda$ can represent decoherence of a state $\rho$. In order to optimise the setup in Fig. \ref{fig:setup} one is required to find $\sup_{\rho}F^Q_{\Lambda(\rho)}$ and the corresponding optimal input state $\rho$. 

The~supremum $F^Q=\sup_{\rho}F^Q_{\Lambda(\rho)}$ of the quantum Fisher information can be expressed by the following \textbf{variational principle}:
\begin{equation}
F^Q=\sup_{X\in L^2(\mathcal{H})}\sup_{|\psi\rangle}\,\langle\psi|\, \Lambda^\dagger\left( -X^2+2i[H,X] \right)   |\psi\rangle,\label{eq:supQF}
\end{equation}
where $|\psi\rangle$ is a normalised vector in $\mathcal{H}$. The set of normalised states can be replaced by the set $S$ of all density matrices. \\

\textbf{Proof.} The order of optimisation is arbitrary.  We have:
\begin{equation}
\sup_{X\in L^2(\mathcal{H})}\,\langle\psi|\, \Lambda^\dagger\left( -X^2+2i[H,X] \right)   |\psi\rangle = \sup_{X\in L^2(\mathcal{H})}\,\mathrm{Tr}\{ \Lambda\left(|\psi\rangle\langle\psi|\right)\, \left( -X^2+2i[H,X] \right) \} .
\end{equation}
We perform differentation w.r.t. coefficients of $X$ in any basis, thus arriving at  the equation $\frac{1}{2}\{X, \Lambda\left(|\psi\rangle\langle\psi|\right)\}=-i[H, \Lambda\left(|\psi\rangle\langle\psi|\right)]$, which is just the equation of the symmetric logarithmic derivative $L_{\Lambda\left(|\psi\rangle\langle\psi|\right)}$. The~matrix $F^{(2)}$ of second-order partial derivatives is negative, because when sandwiched between operators $Y\in L^2(\mathcal{H})$ it gives $Y^*F^{(2)}Y=-2\mathrm{Tr}\{\Lambda\left(|\psi\rangle\langle\psi|\right)Y^2\}$. Hence, this extremum is a maximum.

We are left with $\sup_{|\psi\rangle}\mathrm{Tr}\{\Lambda\left(|\psi\rangle\langle\psi|\right) L_{\Lambda\left(|\psi\rangle\langle\psi|\right) }^2\}=\sup_{|\psi\rangle}F^Q_{\Lambda\left(|\psi\rangle\langle\psi|\right)}$. The convexity of the quantum Fisher information  w.r.t. density matrices implies that $\sup_{|\psi\rangle}F^Q_{\Lambda\left(|\psi\rangle\langle\psi|\right)}=\sup_{\rho}F^Q_{\Lambda(\rho)}= F^Q$. $\qedsymbol$%\\

\section{Algorithm}

%\textbf{Algorithm.} 
Such a variational principle motivates the introduction of the following iterative alternating algorithm to effectively compute the maximum quantum Fisher information $F^Q$. \\

\textbf{Algorithm.} Let $F(\rho,X):=\mathrm{Tr}\{\rho \left( -X^2+2i[H,X] \right) \}$ and $G(X):=-X^2+2i[H,X]$ s.t. $F(\rho,X)=\mathrm{Tr}\{\rho\, G(X)\}$. One starts with the input state $|\psi_0\rangle$ and defines  $\rho_0=\Lambda(|\psi_0\rangle\langle\psi_0|)$.  We know that the maximum value of $F(\rho_0, X)$ is attained for $X=L_{\rho_0}$. The input state $|\psi_1\rangle $  for~the~next step of the procedure is~therefore~chosen to be the eigenvector corresponding to the maximum eigenvalue of the~operator $ \Lambda^\dagger\left(G(L_{\rho_0}) \right) $. 

Analogously, let $|\psi_n\rangle$ be the initial state at n-th step and $\rho_n=\Lambda(|\psi_n\rangle\langle\psi_n|)$. We obtain $|\psi_{n+1}\rangle$ as the eigenvector corresponding to the maximum eigenvalue of the operator $ \Lambda^\dagger\left( G(L_{\rho_n}) \right) $. Therefore the quantum Fisher information of $\rho_n$ increases with $n$:
\begin{equation}
F^Q_{\rho_n}=F(\rho_n, L_{\rho_n})\leq F(\rho_{n+1}, L_{\rho_n})\leq F(\rho_{n+1}, L_{\rho_{n+1}})=F^Q_{\rho_{n+1}}.
\end{equation}

We prove below that in the case of a finite dimension of $\mathcal{H}$ the algorithm provides the maximum quantum Fisher information $\lim_{n\rightarrow\infty}F^Q_{\rho_n}=F^Q$ and~the~optimal input state yield by a subsequence $\lim_{m\rightarrow\infty}|\psi_{n_m}\rangle$.\\

The maximization of the quantum Fisher information is thus achieved by alternatively `moving' along two perpendicular `directions'  $L^2(\mathcal{H})$ and $\Lambda(S)$, where $S$ is the set of density matrices. At~each step we first go as high as possible in `direction' $L^2(\mathcal{H})$ and then as high as possible in `direction' $\Lambda(S)$. Since $F(\Lambda(\rho),X)$ is linear w.r.t. $\rho$, we always arrive at the boundary of $S$ moving in the $\Lambda(S)$ direction, i.e. the chosen $\rho\in S$ is~a~pure state. The algorithm only requires diagonalising two operators:  $\rho_n$ and  $ \Lambda^\dagger\left( G(L_{\rho_n}) \right) $ at each step.\\

%The algorithm can be described as climbing a hill $F$ by only moving along two perpendicular directions $L^2(\mathcal{H})$ and $\Lambda(S)$, where $S$ is the set of density matrices. At~each step we first ascend the hill as high as possible in direction $L^2(\mathcal{H})$ and then as high as possible in direction $\Lambda(S)$. Since $F(\Lambda(\rho),X)$ is linear w.r.t. $\rho$, we always arrive at the boundary of $S$ moving in the $\Lambda(S)$ direction, i.e. the chosen $\rho\in S$ is~a~pure state.\\

\textbf{Proof of convergence.} The increasing sequence $f_n=F(\rho_n,L_{\rho_n})=F^Q_{\rho_n}$ is bounded from above when $H\in L^2(\mathcal{H})$:
%\begin{eqnarray*}
%\sup_{|\psi\rangle}F(\Lambda(|\psi\rangle\langle\psi|),L_{\Lambda(|%\psi\rangle\langle\psi|)})&\leq&\sup_{|\psi\rangle}F(|\psi\rangle\langle\psi|,L_{|%\psi\rangle\langle\psi|})\\&=&\sup_{|\psi\rangle}\left(\langle \psi|H^2|\psi\rangle-\langle %\psi|H|\psi\rangle^2\right)<\infty.
%\end{eqnarray*}
%\begin{equation}
%\sup_{|\psi\rangle}F(\Lambda(|\psi\rangle\langle\psi|),L_{\Lambda(|\psi\rangle\langle\psi|)})\leq\sup_{|\psi\rangle}F(|\psi\rangle\langle\psi|,L_{|\psi\rangle\langle\psi|})=\sup_{|%\psi\rangle}\left(\langle \psi|H^2|\psi\rangle-\langle \psi|H|\psi\rangle^2\right)<\infty.
%\end{equation}
\begin{equation}
\sup_{|\psi\rangle}F^Q_{\Lambda(|\psi\rangle\langle\psi|)}\leq\sup_{\rho} F^Q_{\rho}=\sup_{|\psi\rangle}\left(\langle \psi|H^2|\psi\rangle-\langle \psi|H|\psi\rangle^2\right)<\infty.
\end{equation}
Thus, the $\{f_n\}$ sequence converges to a limit $f^*$. %\\

First, let us prove that $F(\rho,X)=F(\rho,L_\rho)=F(\rho',L_{\rho'})$, where $\rho'=\Lambda(|\psi\rangle\langle\psi|)$ and $|\psi\rangle$ is the eigenvector corresponding to the maximum eigenvalue of $ \Lambda^\dagger\left( G(L_{\rho}) \right) $, implies that $X=L_\rho$, $\rho=\rho'$  and $\rho$ correponds to the~maximum quantum Fisher information, i.e. $F_\rho^Q =F^Q$. 

If $L_{\rho}$ is uniquely defined for $\rho$, the equality $F(\rho,X)=F(\rho,L_\rho)$ implies $X=L_\rho$. On the other hand, the equation $F(\rho,L_\rho)=F(\rho',L_{\rho'})$ leads to $F(\rho,L_\rho)=F(\rho',L_{\rho})$, which, if the maximal eigenvalue eigenspace of $\Lambda^\dagger\left( G(L_{\rho}) \right) $ is~nondegenerate, imposes $\rho=\rho'$. We assume that, for the choice of $L_\rho$ and $|\psi\rangle$ to be unique, these two conditions are fulfilled at all the steps of the algorithm (see~Appendix B).

We prove now that $\rho$ corresponds to the maximum, i.e. $F_\rho^Q =F^Q$:
\begin{eqnarray}
 F\left(\rho+\delta\rho^{(1)} ,L_\rho+\delta X^{(1)}\right)&=&  F\left(\rho+\delta\rho^{(1)} ,L_\rho\right)+\delta\, \mathrm{Tr}\left\{\rho\left( -\{L_\rho,X^{(1)}\}+2i[H,X^{(1)}]\right) \right\}  + O(\delta^2)\nonumber\\
&=&  F\left(\rho+\delta\rho^{(1)} ,L_\rho\right)+ O(\delta^2)
<F(\rho,L_\rho)+ O(\delta^2) ,\label{eq:conv}
\end{eqnarray}
where the second equality is due to  $\{L_\rho,\rho\}=2i[H,\rho]$ and the cyclic property of the trace, and the inequality follows from $\rho$ being the optimal state in $\Lambda(S)$ for $X=L_\rho$ ($\rho=\rho'$). Since $F(\rho,X)$ is concave w.r.t. $(\rho,X)$, Eq.~(\ref{eq:conv}) implies that $F(\rho,L_\rho)$ is the global maximum $F^Q$.\\

If the algorithm gets stuck, i.e. $f_n=F(\rho_n,L_{\rho_n})$=$F(\rho_{n+1},L_{\rho_{n+1}})=f_{n+1}$, we have arrived at~the~maximum $F^Q$, thus ending the proof for this case.

Let us consider the opposite case in which $f_n<f_{n+1}$ for all $n\in\mathbb{N}$. We assume that the dimension of $H$ is finite from now on. As $\rho_n$ are elements of $S$, which is compact, one can choose a convergent subsequence $\{\rho_{n_m}\}_{m\in\mathbb{N}}$; let $\rho^*$ denote the limit of this subsequence. 
%The set $B(0,||H||_{HS})$ is also compact, therefore, again, one can choose a subsequence of $\{\rho_{n_m}\}_{m\in\mathbb{N}}$ s.t. the corresponding sequence of symmetric logarithmic derivatives converges to $L^*$. Let us, for the sake of simplicity, keep $\{\rho_{n_m}\}_{m\in\mathbb{N}}$ to denote the latter subsequence.% 
We assume that $L_\rho$ is continuous w.r.t. $\rho$, then, together with the continuity of $F(\Lambda(\rho),X)$ w.r.t. $(\rho,X)$, it implies $F(\rho^*,L_{\rho^*})=\lim_{m\rightarrow\infty} F(\rho_{n_m},L_{\rho_{n_m}})=f^*$. 

The eigenvector $|\psi^*\rangle$ corresponding to the maximum eigenvalue of $ \Lambda^\dagger\left( G(L_{\rho^*}) \right)$ leads to $\rho'=\Lambda(|\psi^*\rangle\langle\psi^*|)$ with the same quantum Fisher information $F_{\rho'}^Q=f^*$ as $\rho'$ is the limit of the subsequence $\{\rho_{n_m+1}\}_{m\in\mathbb{N}}$. Therefore, if $L_{\rho^*}$ is uniquely defined and $\Lambda^\dagger(G(L_{\rho^*})$ has a nondegenarate maximal eigenvalue eigenspace, we obtain $f^*=F(\rho^*,L_{\rho^*})=F^Q$  according to the first part of the proof.

Therefore, we have proved the convergence to the maximum, i.e. $f^*=F^Q$, as~long~as, at each step, the choice of $|\psi_n\rangle$ and $L_{\rho_n}$ is unique (see Appendix B). $\qedsymbol$\\

%Let us now assume that $f_n<f_{n+1}$ for all $n\in\mathbb{N}$ and, by contradiction,  $f^*<F^Q$. The set $C_{\epsilon}=\{(\rho,X)\in S\times B(0,||H||_{HS}):\,  f^*-\epsilon\leq F(\Lambda(\rho),X)\leq f^*\} $ is compact as $F(\Lambda(\rho),X)$ is a continuous function of $(\rho,X)$ and $S\times B(0,||H||_{HS})$ is compact. Thus, the continuous function $\Delta F(\rho,X):= F(\rho',L_{\rho'})-F(\rho,X)$ attains its minimal value $\delta_{\epsilon}$ on the compact $C_{\epsilon}$ and $\delta_{\epsilon}>0$ according to the first part of the proof as $f^*<F^Q$. From the convergence $\lim_{n\rightarrow\infty}f_n=f^*$ we have that $f^*-\epsilon\leq f_n\leq f^*$ for a sufficiently large $n$, but then $f_{m+1}-f_m=\Delta F(\rho_m,L_{\rho_m})\geq\delta_\epsilon>0$ for $m\geq n$. This implies that for a sufficienly large $m$ we have $f_{m}>f^*$ leading to a contradiction since $f^*=\lim_{n\rightarrow\infty}f_n$ and $f_n$ increases with $n$. 

The above proof is closely related to Chapter 10.3 in~\cite{RWY}, where proof of convergence is presented for an alternating algorithm in the case of $\sup_{u_1\in A_1} \sup_{u_2\in A_2} f(u_1,u_2)$, where $f$ is a strictly concave real-valued function and $A_i$ is a~compact and convex subset of $\mathbb{R}^{n_i}$, $i=1,2$. In our case the function $F(\Lambda(\rho),X)$ is linear w.r.t. $\rho$ and strictly concave w.r.t. $X$, and, thanks to the~definition of $L_\rho$, the $L^2(\mathcal{H})$ space in the variational principle in Eq.~(\ref{eq:supQF}) can be restricted to~the~compact and convex set $B(0,||H||_{HS})$ of operators with Hilbert-Schmidt norm not greater than $||H||_{HS}$ as $||L_\rho||_{HS}\leq||H||_{HS}$.

\section{COMMENTS}

A variational principle  analogous to Eq.~(\ref{eq:supQF}) holds for the maximum Fisher information $F_{\varphi,\{\Pi_x\}}$ for a fixed POVM measurement  $\{\Pi_x\}_{x\in X}$ :
\begin{equation}
F_{\varphi,\{\Pi_x\}}=\sup_{D\in L^2(X)}\sup_{|\psi\rangle}\,\langle\psi|\, \Lambda_\varphi^\dagger\left( -X^D_2+2i[H,X^D_1] \right)   |\psi\rangle, \label{eq:supF}
\end{equation}
where $X^D_j=\int_X\mathrm{d}x \,D(x)^j\,\Pi_x $, $\Lambda_{\varphi}(\rho)= e^{-i \varphi H }\Lambda(\rho)e^{i \varphi H }$, $L^2(X)$ is the set of square-integrable real-valued functions on~$X$ and $|\psi\rangle$ is a normalised vector in $\mathcal{H}$. If the measurement is a projective von Neumann measurement, $X_2=X_1^2$. 

The proof of Eq.~(\ref{eq:supF}) is presented in Appendix C. Here, let us simply state that for a given $|\psi\rangle$ the supremum $\sup_{D\in L^2(X)} \,\langle\psi|\, \Lambda_{\varphi}^\dagger\left( -X_2+2i[H,X_1] \right)   |\psi\rangle$  is attained at:
\begin{equation}
 D_L(x)=\left\{ \begin{array}{ll}
\frac{\mathrm{Tr}\{-i [H,\Lambda_{\varphi}\left(|\psi\rangle\langle\psi|\right)]\Pi_x\}}{\mathrm{Tr}\{\Pi_x\Lambda_{\varphi}\left(|\psi\rangle\langle\psi|\right)\}}, & \textrm{if\quad} \mathrm{Tr}\{\Lambda_{\varphi}\left(|\psi\rangle\langle\psi|\right)\Pi_x\}>0,\\
0, & \textrm{elsewhere,}
\end{array} \right. \label{eq:D}
\end{equation}
and the supremum value equals $\langle\psi|\, \Lambda_{\varphi}^\dagger( X^{D_L}_2)  |\psi\rangle$.

Using Eq.~(\ref{eq:D}) one can introduce an analogous alternating iterative algorithm to optimise the Fisher information for a fixed POVM measurement.\\

It can be easily demonstrated that the parabolic function $G(X)=\left( -X^2+2i[H,X] \right)$ appearing in Eq.~(\ref{eq:supQF}) and  in $F(\rho,X)=\mathrm{Tr}\{\rho \, G(X)\}$ can be replaced by any~other function $G':\, L^2(\mathcal{H})\rightarrow \mathcal{B}(\mathcal{H})$ s.t. $\mathrm{Tr}\{ \rho \,G'( X)\}$   has a global maximum w.r.t $X$ equal to $\mathrm{Tr}\{ \rho \,X_\rho^2\}$, where $\frac{1}{2}\{X_\rho, \rho\}=-i[H, \rho]$.\\

One can consider a  \textbf{general case} when the parameter $\varphi$ is encoded on a state $\rho$ via a channel $\Lambda_\varphi$, i.e. $\rho_\varphi=\Lambda_\varphi(\rho)$. This decribes e.g. decoherence channels which do not commute with $H$: $\Lambda_\varphi (\rho)\neq e^{-i \varphi H}\Lambda_0(\rho) e^{i \varphi H}$. Let $\Lambda_\varphi'$ denote $\frac{\mathrm{d}}{\mathrm{d}\varphi}\Lambda_\varphi$. We obtain the following variational principle:
\begin{equation}
F^Q=\sup_{X\in L^2(\mathcal{H})}\sup_{|\psi\rangle}\,\langle\psi|\, \left(-\Lambda_\varphi^\dagger(X^2)+2\Lambda_\varphi'^\dagger(X) \right)   |\psi\rangle,\label{eq:supQFgen}
\end{equation}
where $|\psi\rangle$ is a normalised vector in $\mathcal{H}$. The set of normalised states can be replaced by the set of density matrices. 

The general variational principle in Eq.~(\ref{eq:supQFgen}) suggests introducing an alternating iterative algorithm analogous to the one discussed above. In the case of a fixed POVM measurement Eq.~(\ref{eq:supQFgen}) can also easily be modified.\\

If we restrict the set of normalised states, or equally, the set of density matrices in Eq.~(\ref{eq:supQF}) to a \textbf{class of states}, e.g. MPS states or gaussian states, we obtain a variational principle for the maximum quantum Fisher information w.r.t. this class $C$:
\begin{equation}
\sup_{D\in L^2(X)}\sup_{\rho\in C}\ F(\rho,X)=\sup_{\rho\in C}\sup_{X\in L^2(\mathcal{H})}F(\rho,X)=\sup_{\rho\in C}F_\rho^Q.
\end{equation}
Furthermore, if $C$ is convex, an alternating iterative algorithm can be used.  All this is also true for the variational principles in Eq.~(\ref{eq:supF}) and (\ref{eq:supQFgen}). 
\\

%Analogous variational princpile can be introduced for the Fisher information in the case of  multiparameter estimation. \\

Initially, we obtained the variational principle in Eq. (\ref{eq:supQF}) using Bayesian estimation in the limit of a deterministic prior distribution, which is explained in Appendix A.

\section*{ACKNOWLEDGEMENTS} 

The author is grateful to Rafa\l  \,\,Demkowicz-Dobrza\'nski for fruiftul discussions on Bayesian estimation and Marcin Jarzyna for~checking the efficient performance of the algorithm. 
%for several quantum channels in the case of the $H$ generator which describes a two-arm interferometer.  
Helpful comments on this draft by M\u{a}d\u{a}lin Gu\c{t}\u{a} and Luis A. Correa are gratefully acknowledged. This research was supported by Polish NCBiR under the ERA-NET CHIST-ERA project QUASAR.%\newpage

\begin{appendix}

\section{Fisher information in Bayesian estimation}

For the Bayesian estimation with a prior distribution $g(\varphi)$, we have an interesting connection with the Fisher information in the case of the Gaussian $g$. Let us assume, without loss of generality, that the mean of this distribution equals 0. Let $\Delta^2_{prior}$ denote the prior variance $\int  \mathrm{d}\varphi\,g({\varphi})\,\varphi^2$.

The average variance of an estimator $\widetilde{\varphi}$ is defined as:
\begin{equation}
\Delta^2\widetilde{\varphi}=\int  \mathrm{d}\varphi\,g({\varphi}) \int \mathrm{d}x\, p_{\varphi}(x)\, \left(\widetilde{\varphi}(x)-\varphi\right)^2.
\end{equation}
In the case of Bayesian estimation the best estimator is known to be the conditional expected value, defined as follows:
\begin{equation}
\widetilde{\varphi}(x)=\left\{ \begin{array}{ll}
\frac{\int\mathrm{d}\varphi\,g({\varphi})\, p_{\varphi}(x)\,\varphi}{\int\mathrm{d}\varphi\,g({\varphi})\, p_{\varphi}(x)}, & \textrm{if\quad} \int\mathrm{d}\varphi\,g({\varphi})\, p_{\varphi}(x)>0,\\
0, & \textrm{elsewhere.}
\end{array} \right. \label{eq:bestBayesian}
\end{equation}
We have:
\begin{equation}
\Delta^2\widetilde{\varphi}=\Delta^2_{prior}- \int  \mathrm{d}\varphi\,g({\varphi})\int \mathrm{d}x\, p_{\varphi}(x)\, \widetilde{\varphi}(x)^2.
\end{equation}
For the gaussian distribution $g(\varphi)\propto \exp(-\frac{\varphi^2}{2\Delta^2_{prior}})$ we have $g(\varphi) \,\varphi=\Delta^{-2}_{prior} \frac{\mathrm{d} \,g({\varphi})}{\mathrm{d}\varphi} $. Thus:
\begin{eqnarray}
\Delta^{-2}_{prior}\,\widetilde{\varphi}(x)&=& \frac{\int\mathrm{d}\varphi  \frac{\mathrm{d} \,g({\varphi})}{\mathrm{d}\varphi} \, p_{\varphi}(x)}{\int\mathrm{d}\varphi\,g({\varphi})\, p_{\varphi}(x)} 
=- \frac{\int\mathrm{d}\varphi \,g({\varphi}) \frac{\mathrm{d} \, p_{\varphi}(x)}{\mathrm{d}\varphi}}{\int\mathrm{d}\varphi\,g({\varphi}) p_{\varphi}(x)} \textrm{\quad and}\\
\frac{1-\frac{\Delta^2\widetilde{\varphi}}{\Delta^2_{prior}}}{\Delta^2_{prior}}&=&\frac{ \int  \mathrm{d}\varphi\,g({\varphi})\int \mathrm{d}x\, p_{\varphi}(x)\, \widetilde{\varphi}(x)^2}{\Delta^4_{prior}}\nonumber\\
&=&  \int \mathrm{d}x \int\mathrm{d}\varphi\,g({\varphi}) p_{\varphi}(x)\, \left( \frac{\int\mathrm{d}\varphi \,g({\varphi}) \frac{\mathrm{d} \, p_{\varphi}(x)}{\mathrm{d}\varphi}}{\int\mathrm{d}\varphi\,g({\varphi}) p_{\varphi}(x)}\right )^2 = F^g_0, \label{eq:FisherBayes}
\end{eqnarray}
where $ F^g_0$ is the Fisher information at $\phi=0$ for a new family of distributions $p^{g}_{\phi}(x)= \int\mathrm{d}\varphi\,g({\varphi}) p_{\varphi+\phi}(x)$. If we choose from different families of distributions $\{p_{\varphi}\}$, e.g. various initial states $\rho$, minimisation of $\Delta^2\widetilde{\varphi}$ is equivalent to maximisation of $F^{g}_0$. 

In the limit $g(\varphi)\rightarrow\delta(\varphi)$:
\begin{equation}
\lim_{\Delta_{prior}\rightarrow 0} F^g_0= \int \mathrm{d}x  \,p_{0}(x)\, \left( \frac{ \frac{\mathrm{d} \, p_{\varphi}(x)}{\mathrm{d}\varphi}|_{\varphi=0}}{p_{0}(x)}\right )^2=F_0.
\end{equation}%\\

%ABOVE INSTEAD OF THE MAIN TEXT (If IN THE MAIN TEXT)

For the setup described in the main text we have $\rho_{\varphi}=e^{-i \varphi H}\rho e^{i \varphi H}$ and $p_{\varphi}(x)=\mathrm{Tr}\{\Pi_x\rho_{\varphi}\}$, where $\rho=\Lambda(|\psi\rangle\langle\psi|)$ is the initial state and $\{\Pi_x\}_{x\in X}$ describes the measurement. For a prior distribution $g$ we look for a choice of $\rho$, $\{\Pi_x\}_{x\in X}$  and $\widetilde{\varphi}$ which minimises $\Delta^2\widetilde{\varphi}$. The case of $\varphi$ being the frequency difference in an atomic clock was discussed in~\cite{KM}, where the relation in Eq.~(\ref{eq:FisherBayes}) above is already reffered to in Eq.~(6). In the case of a fixed measurement, let~$F_{\rho}^g$ denote $F_0^g$ for the initial state $\rho$.

To optimise the setup with a fixed measurement, we perform:
\begin{equation}
 \sup_{|\psi\rangle} F_{\Lambda(|\psi\rangle\langle\psi|)}^g=\sup_{|\psi\rangle} \sup_{\tilde{\varphi}} \frac{1-\frac{\Delta^2\widetilde{\varphi}}{\Delta^2_{prior}}}{\Delta^2_{prior}}.
\end{equation}
By exchanging the suprema and taking the limit $\Delta_{prior}\rightarrow 0$ we arrive at the variational principle for $F_{\{\Pi_x\}}$ in Eq.~(\ref{eq:supF}). 
A function $D^L$ is the limit of best Bayesian estimators for $\rho=\Lambda(|\psi\rangle\langle\psi|)$. 

When the measurement is also being optimised, we arrive at Eq.~(\ref{eq:supQF}) and the symmetric logarithmic derivative of~$\Lambda(|\psi\rangle\langle\psi|)$ is the limit of the operators $\int_X\mathrm{d}x\,\widetilde{\varphi}(x)\,\Pi_x$ encoding the best Bayesian estimators and projective measurements for this state.\\

In~\cite{RDD} an alternating iterative algorithm was used to optimise the initial states and measurements in phase estimation on the $[-\pi,\pi]$ interval with a given prior distribution and a sinusoidal cost function.  In~\cite{KM} an analogous algorithm was introduced for frequency estimation with a given prior distribution and a square cost function. In this draft an alternating iterative algorithm, to optimise the Fisher information, in both cases of a measurement being fixed or~capable of being optimised, is both proposed and proved. 

\section{Unique choices in the algorithm}

As stated above, the unique choices of $|\psi_n\rangle$ and $L_{\rho_n}$ at each step are cruicial.

Let us note that $L_{\rho}$ is uniquely defined only when $\rho$ has a maximum possible rank. Let $\rho=\sum_{j=1}^{d'}\lambda_j\,|\lambda_j\rangle\langle\lambda_j|$, where $d'$ is the rank of $\rho$ and $\{|\lambda_j\rangle\}_{j=1}^{d'}$ are orthonormal vectors in $\mathcal{H}$. It can be shown that the equation $\frac{1}{2}\{X_\rho, \rho\}=-i[H, \rho]$ imposes the following form of $L_\rho$:
\begin{equation}
L_\rho=\sum_{j,k=1}^{d'} \left(  \frac{\lambda_k-\lambda_j }{\lambda_k+ \lambda_j }\langle\lambda_k| i H|\lambda_j\rangle\,\right) \, -\,  P_{V^\perp}i H P_{V} +  P_{V}i H P_{V^\perp}  \, +\,L_\rho^a \,. \label{eq:Lrho}
\end{equation}
where $V^{\perp}$ is the orthogonal complement of $V$, $P_V$, $P_{V^\perp}$ are the orthogonal projections on $V$ and $V^{\perp}$ respectively and $L_\rho^a:\,V^{\perp}\rightarrow V^{\perp}$ represents the part of $L_\rho$ which can be defined arbitrarily. From Eq.~(\ref{eq:Lrho}) we see that $L_\rho$ is only uniquely defined when $V=\mathcal{H}$. 

If $ L_\rho$  is not uniquely defined, we set $L_{\rho}^a=0$ in the algorithm. In the case of a unitary channel $\Lambda$,  the state $\Lambda(|\psi\rangle\langle\psi|)$ has rank equal to $1$. Nevertheless, we observed numerically that as long as the initial state $|\psi_0\rangle$ has non-zero expansion coefficients in the eigenbasis of $H$, the algorithm converges to the maximum value $F^Q=\sup_{|\psi\rangle}\left(\langle \psi|H^2|\psi\rangle-\langle \psi|H|\psi\rangle^2\right)$. This condition is related to the fact that when $\rho_n$ has a block diagonal form in the eigenbasis of $H$, $L_{\rho_n}$ also has an analogous block diagonal form according to Eq.~(\ref{eq:Lrho}) and, moreover, this form is preserved at all the following steps of the algorithm.  Assuming the nondegeneracy of the maximal eigenvalue of $\Lambda^\dagger(L_{\rho_n})$, $\rho_{n+1}$ is only supported on one of the blocks, $L_{\rho_{n+1}}$ is not uniquely defined and, from this point on, the algorithm is effectively restricted to the subspace of $\mathcal{H}$ related to this block. Therefore, in such a case the algorithm will fail to provide the maximum quantum Fisher information $F^Q$, unless that maximum corresponds to the state which is supported only on this block. 

In the case of a nonunitary $\Lambda$ which commutes with $H$, i.e. $\Lambda (e^{-i \varphi H}\rho e^{i \varphi H})=e^{-i \varphi H}\Lambda(\rho) e^{i \varphi H}$, a block diagonal form of $\rho_n$ in the eigenbasis of $H$ is also preserved by the algorithm. Therefore, in both these cases we make the following \textbf{proposition}.\\

If at each step of the algorithm $\rho_n$ is irreducible w.r.t. the direct sums of the eigenspaces of $H$, the algorithm will converge to the maximum value $F^Q$. \\

By saying that $\rho$ is irreducible w.r.t. the direct sums of the eigenspaces of $H$, we understand the following: if for a subspace $V\subset\mathcal{H}$ we have both $\rho V\subset V$ and $H V\subset V$, then either $V=\{0\}$ or $V=\mathcal{H}$. This is equivalent to saying that $\rho$ has a block diagonal form in the eigenbasis of $H$ which consists just of one block.

%Randomness can be introdued!

%\textbf{Symmetries.}  In the presence of a symmetry, one may need to define a way of choosing  representatives of the~equivalence classes implied by this symmetry. Let us discuss the example of $\Lambda$ commuting with $H$, i.e. $\Lambda (e^{-i \varphi H}\rho e^{i \varphi H})=e^{-i \varphi H}\Lambda(\rho) e^{i \varphi H}$. The global phases of the eigenbasis of the $H$ generator are arbitrary. As a result, all the states $|\psi\rangle$ whose expansion coefficients in this basis have the same modulus, correspond to the same quantum Fisher information. The vector coresponding to the maximum $F^Q$, among others, is not uniquely defined. At the beginning of each step one can modify all the coefficients of $|\psi_n\rangle$ to be positive in the eigenbasis of $H$ so that, as long as the maximal eigenvalue eigenspace of $ \Lambda^\dagger\left( G(L_{\rho_{n-1}}) \right)$ is nondegenerate, the choice will be unique.

%NAPISZ ZE MOZE NIEKONIECZNIE!!!!

\section{Proof of the variational principle in the case of a fixed measurement}

Let $F_{\Lambda_\varphi(|\psi \rangle \langle\psi|),\{\Pi_x\}}$ be the~Fisher information for a state $\Lambda_\varphi(|\psi \rangle \langle\psi|)$ and a measurement $\{\Pi_x\}_x\in X$. It is enough to prove that for a normalised vector $|\psi\rangle$ in $\mathcal{H}$:
\begin{equation} 
\sup_{D\in L^2(X)}\langle\psi| \Lambda_\varphi^\dagger\left( -X^D_2+2i[H,X^D_1] \right)  |\psi\rangle = \langle\psi| \Lambda_\varphi^\dagger\left( -X^{D_L}_2+2i[H,X^{D_L}_1] \right)   |\psi\rangle , \label{eq:apI1}
\end{equation}
where $D_L$ is defined in Eq.~(\ref{eq:D}), and that
\begin{equation} 
F_{\Lambda_\varphi(|\psi \rangle \langle\psi|),\{\Pi_x\}}= \langle\psi| \Lambda_\varphi^\dagger\left( -X^{D_L}_2+2i[H,X^{D_L}_1] \right)   |\psi\rangle. \label{eq:apI2}
\end{equation}
Once more, the convexity of the Fisher information w.r.t. a density matrix, guarantees  $F_{\varphi,\{\Pi_x\}}=\sup_{|\psi\rangle} F_{\{\Pi_x\}, \Lambda_{\varphi}(|\psi \rangle \langle\psi|)}$.\\

The equality in Eq.~(\ref{eq:apI1}) is proved by taking either the derivative w.r.t. $D(x)$ when $X$ is discrete, or a functional derivative w.r.t. to $D(x)$ in a continuous case. $D_L(x)$ can be defined arbitrarily when $\mathrm{Tr}\{\Pi_x\Lambda_\varphi\left(|\psi\rangle\langle\psi|\right)\}=0$. 

In Eq.~(\ref{eq:apI2}), the identity is a consequence of the definition of the $D_L$ function:
\begin{eqnarray} 
 \langle\psi| \Lambda_\varphi^\dagger\left( -X^{D_L}_2+2i[H,X^{D_L}_1] \right)   |\psi\rangle&=&\mathrm{Tr}\{  \Lambda_\varphi(|\psi\rangle\langle\psi| ) X^{D_L}_2 \} \nonumber\\
&=&\int_{X_{\Lambda_\varphi}}\mathrm{d}x \,\mathrm{Tr}\{\Pi_x\Lambda_\varphi\left(|\psi\rangle\langle\psi|\right)\} \left(\frac{\mathrm{Tr}\{-i[H,\Lambda_\varphi\left(|\psi\rangle\langle\psi|\right)]\Pi_x\}}{\mathrm{Tr}\{\Lambda_\varphi\left(|\psi\rangle\langle\psi|\right)\Pi_x\}}\right)^2  \nonumber\\
&=&\int_{\{x:\, p_\varphi(x)\neq 0\}}\mathrm{d}x\, p_\varphi(x) \,\left(\frac{\frac{\partial p_{\varphi}(x)}{\partial \varphi}}{p_\varphi(x)}\right)^2 = F_{{\Lambda_\varphi\left(|\psi\rangle\langle\psi|\right),\{\Pi_x\}}}
\end{eqnarray}
where  $X_{\Lambda_\varphi}=\{x:\,\mathrm{Tr}\{\Lambda_\varphi\left(|\psi\rangle\langle\psi|\right)\Pi_x\}\neq0\}$ and $p_\varphi(x)=\mathrm{Tr}\{\Lambda_\varphi\left(|\psi\rangle\langle\psi|\right)\Pi_x \}$. $\qedsymbol$

\end{appendix}

\end{document}